\documentclass[review]{elsarticle}

\usepackage{lineno}
\usepackage{amsmath}
\usepackage{subfig}
\usepackage[toc,page]{appendix}
\usepackage{blindtext}
\usepackage{natbib}

\usepackage{relsize}
\usepackage{latexsym}
\usepackage{graphicx}
\usepackage{siunitx}
\usepackage{booktabs}
\usepackage{comment}
\usepackage{makecell}
\usepackage{relsize}
\usepackage{array,multirow,tabularx}

\usepackage[utf8]{inputenc}
\usepackage[T1]{fontenc}
\usepackage{lmodern}
\usepackage[figurename=Fig.,labelfont=bf,labelsep=period]{caption}
\usepackage{newtxtext,newtxmath}
\usepackage[citecolor=black,linkcolor=black]{hyperref}
\usepackage{tablefootnote}

\modulolinenumbers[5]

\journal{Journal of the Mechanics and Physics of Solids}

\biboptions{authoryear}

\begin{document}

\begin{frontmatter}

\title{Physics-based Constitutive Equation for Thermo-Chemical Aging in Elastomers based on Crosslink Density Evolution} 


\author{Maryam Shakiba\corref{mycorrespondingauthor}$^{1}$}
\cortext[mycorrespondingauthor]{Corresponding author.}
\ead{mshakiba@vt.edu}
\author{Aimane Najmeddine$^{1}$}
\address{$^{1}$Department Civil and Environmental Engineering, Virginia Tech, USA}
\begin{abstract}
This paper presents a physics-based constitutive equation to predict the thermo-chemical aging in elastomers. High-temperature oxidation in elastomers is a complex phenomenon. The macromolecular network of elastomers' microstructures undergoes chain scission and crosslinking under high temperature and oxygen diffusion conditions. In this work, we modify the network stiffness and the chain extensibility in the Arruda-Boyce well-known eight-chain constitutive equation to incorporate the additional Helmholtz free energy due to network changes in elastomers' microstructures. The effect of network evolution due to aging in changing the shear modulus and the number of chain monomers is considered. The modification is based on chemical characterization tests, namely the equilibrium swelling experiment to measure the crosslink density evolution. The developed constitutive equation predicts the mechanical responses of thermo-chemically aged elastomers independent of any mechanical tests on aged samples. The proposed constitutive equation is validated with respect to a comprehensive set of experimental data available in the literature that were designed to capture thermo-chemical aging effects in elastomers. The comparison showed that the constitutive equation can accurately predict the intermittent tensile tests based on crosslink density evolution input. The developed constitutive equation is physics-based, simple, and includes minimal material parameters.
\end{abstract}


\end{frontmatter}



\section{Introduction}
Elastomers are widely used in a variety of engineering applications such as medical devices, aerospace components, and civil infrastructure thanks to their exceptional mechanical and chemical properties. Under operations, elastomers often experience mechanical loading and are exposed to oxygen and elevated temperatures. Thermo-chemical aging due to oxygen diffusion or thermal stresses is an irreversible mechanism that degrades the mechanical properties of elastomers and compromises their desired long-term performance \citep{HUTCHINSON1995703}. The durability of elastomers is, therefore, strongly dependent on thermo-chemical aging coupled with mechanical loading. Thus, understanding and predicting of degradation of elastomers' mechanical properties under thermo-chemical aging play a crucial role in product development.

The mechanical behavior of elastomers is intrinsically multiplex. Aging as multiphysics thermo-chemo-mechanical phenomena occurs and increases to the level of complexity. Thermo-chemical aging phenomena happen through changes in the macromolecular network such as chain scission, crosslink formation, crosslink breakage under high temperature or radiation, and transformation of linkages (e.g., \cite{Blum1951,GILLEN1995149,COLIN2004309,SHAW20052758,Budzien2008,petrikova2011influence,spreckels2012investigations,COQUILLAT20071326,COLIN2007886,Wineman2009}). Thermo-chemical aging in elastomers often leads to material embrittlement \citep{verdu2012oxydative}. The fundamental understanding of the thermo-chemo-mechanical aging of elastomers is challenging due to the fact that chain scission and crosslinking are both at work, the process is slow, and displacements are large. 

Well-established mechanical tests (i.e., the continuous and intermittent relaxation tests) were developed to detect the effects of thermo-chemical aging on the mechanical responses of elastomers as a function of temperature and oxygen exposure in various media \citep{Tobolsky1944,Andrews1946,Scalan1958,Dunn1959}. The mechanical characterization for engineering purposes is usually limited to stress and strain analysis, which is not enough to describe the complicated changes in elastomers' macromolecular network. Moreover, chemical changes of elastomer under aging conditions have been studied by several researchers \citep{CELINA2000171,CELINA2005395,NAGLE20071446,steinke2011numerical,Saux20120901}. \citet{celina2013review} presented a comprehensive review on the approaches for predicting elastomers thermo-chemical aging.

During the last decades, several researchers modeled thermo-oxidative aging of elastomers and polymers including chemical reactions, diffusion, and mechanical coupling (e.g., \cite{achenbach2003finite,SHAW20052758,pochiraju2006modeling,GIGLIOTTI2011431, steinke2011numerical,johlitz2011chemical,Johlitz2013,JOHLITZ2014138,Lejeunes2018constitutive,KONICA2020103858}). Phenomenological and thermodynamic-based frameworks were developed to combine diffusion and reaction expressions to link mechanical responses to chemical kinetics (e.g., \cite{WISE19971929,WISE1997565,pochiraju2006modeling}). Johlitz et al. \citep{LION20121227,JOHLITZ2014138,Johlitz2013} formulated a constitutive approach by evaluating the Clausius-Planck inequality and continuum damage mechanics. They employed the concept of state variables to describe the scission reactions of the primary network and the crosslink reactions creating the secondary network. \citet{WinemanShaw2019} developed a constitutive theory for elastomers at elevated temperatures representing the chemical kinetics of scission, re-coiling, and re-crosslinking, that were expressed in terms of activation energies.


Moreover, micro-mechanical constitutive equations based on statistical mechanics of polymer structure were developed (e.g., \cite{MOHAMMADI2020109108,Beurle2020,KONICA2021104347}). \citet{MOHAMMADI2020109108} developed a micro-mechanical model based on the competition between chain-scission and crosslinking events occurring at the polymer network during oxidation. \citet{KONICA2021104347} developed an oxidation reaction-informed evolving network theory to connect the microscale network evolution with macroscopic damage occurring in polymers. They used the transient network theory based on the statistical mechanics framework of the polymer chains to model the microscale network evolution yielded by the chemical reactions. \citet{ZHI201915} established a hyper-viscoelastic constitutive equation according to the alteration of the macromolecular network observed in experiments, involving the micro-mechanical deformation mechanisms to capture the mechanical behavior of Styrene-Butadiene Rubber (SBR) at different aging states.

In summary, the developed constitutive equations have either assumed the form of thermodynamic energies phenomenologically, been dependent on the types of mechanical tests, or provided no robust link of the changes in the elastomers' macromolecular network to chemical pathways. Most of the more complicated constitutive equations are essentially mechanical theories and have yet to be verified fully experimentally. Moreover, the more recently developed constitutive equations in the literature despite making great progress rely on several either mechanical tests or chemistry kinetics characterizations to obtain numerous model parameters. Thus, although much work has been accomplished in the experimental and the simulation aspects, the link between the network evolution and the mechanical responses in elastomers' aging is still missing. 

This paper will contribute to the missing relationship between the chemical macromolecular changes and the mechanical responses of elastomeric materials upon thermo-chemical aging. This work develops a physics-based and thermodynamically consistent constitutive equation for elastomers thermo-chemical aging coupled with the mechanical responses. The main contribution is to connect the thermodynamic-based formulation and the form of stored energy directly to chemical characterization experiments. The Helmholtz free energy is modified to include extra stored energy in the material due to the consequent macromolecular network evolution upon aging. The modification is based on chemical characterization tests, namely the equilibrium swelling experiment, to measure the crosslink density evolution. Thus, we obtain the additional stored energy based on chemical characterization tests. The developed constitutive equation can predict the mechanical responses of thermo-chemically aged elastomers independent of mechanical tests on aged specimens. The paper is organized as follows. The thermo-chemical aging phenomena and the hypothesis to develop the constitutive equation are presented in Section~\ref{cont model development}. Section~\ref{cont model validation} presents the validation of the developed constitutive equations versus several sets of experimental data available in the literature.


\section{Constitutive relationship for thermo-chemical aging in elastomers} \label{cont model development}
This section first explains the thermo-chemical aging phenomenon in elastomers due to high-temperature conditions and oxygen diffusion. Based on the fundamental understanding, a physics-based constitutive equation is proposed, which uses the evolution of crosslink density due to thermo-chemical aging to acquire the stored energy in elastomers. 

\subsection{The thermo-chemical aging mechanism} \label{subsec: thermo-chemical aging mechanism}
Thermo-chemical aging in elastomers due to high-temperature conditions and oxygen diffusion causes chemical reactions which lead to chain scission, crosslink breakage, and crosslink formation (e.g., \cite{Blum1951,GILLEN1995149,COLIN2004309,SHAW20052758,Budzien2008,petrikova2011influence,spreckels2012investigations,COQUILLAT20071326,COLIN2007886}). Chain-scission reactions break the chains and cause degradation of the original network. Meanwhile, more crosslinks form between the original polymer chains as well as the newly formed chains due to chain-scission. It has been argued in the literature that thermo-chemical aging can be described by two competing mechanisms, the network degradation and the network reformation process (e.g., \cite{achenbach2003finite,Wineman2009,MOHAMMADI2020109108,Beurle2020}). The relative rate of chain-scission and crosslink formation essentially determines if the material becomes more ductile or brittle. Although it must be noted that even if the thermo-chemical kinetics of the chain-scission and crosslinking reactions are similar, the aged material would not have the same toughness like that of the virgin material due to changes in the length of polymer chains \citep{MOHAMMADI2020109108}.

The literature agrees that most polymer chains tend to crosslink under oxidative conditions, leading to an increase in the modulus and the hardening with embrittlement \citep{WISE1995403,WISE19971929,WISE1997565,CELINA1998493,Celina2000,CELINA2000171,celina2013review}. \citet{hamed1999tensile} performed a series of swelling measurements on gum and carbon black filled SBR and Natural Rubber (NR) materials and showed that immediately after an initial aging period, the elastomers exhibit an increase in the crosslink density. The authors then concluded that after an initial aging duration where network chain formation and disruption events are similar, the network formation occurs in greater amounts compared to the disruption when aging time is further increased. \citet{SHAW20052758} and \citet{JOHLITZ2014138} looked more closely at the thermo-chemical aging phenomena in elastomers and observed that the increase in the elastomer stiffness is more pronounced than the network degradation. Tensile tests conducted on several aged specimens showed an increase in the stiffness even under low temperatures and short aging time. However, the difference in the relaxation test data (i.e., tests for which the stretched specimens are being aged and the stress is continuously measured) was less at lower temperatures. Therefore, the authors concluded that network reformation is more dominant in thermo-chemical aging of elastomers \citep{SHAW20052758,JOHLITZ2014138}. Similarly, \citet{Abdelaziz2021} conducted several mechanical experiments on aged SBR for different temperatures and exposure times. They observed higher stiffness at higher temperatures (temperatures up to $100^{\circ}C$) and at longer exposure times and concluded that such behavior is an indicator of crosslinking as the predominant mechanism of thermo-chemical aging for SBR material. Moreover, based on a few chemical characterization tests, \citet{ZHI201915} investigated the elastomer's chain network during aging and showed that chain-scission reactions remain nearly unchanged in the initial aging stage; however, as aging time increases, more dangling chain ends are formed and crosslink density increases. 
Finally, \citet{KONICA2021104347} stated that an oxidized product has a more crosslinked network of smaller chains compared to the unaged polymer with the longer crosslinked network. 

Based on these obtained fundamental understanding of the physical and chemical changes in elastomers network under thermo-chemical aging, in this work, we neglect the dissipative energies due to chain-scission and crosslink breakage. We assume that extra energy is being stored in elastomers due to the formation of crosslinks under high temperature and oxygen diffusion conditions. The proposed physics-based constitutive equation obtained based on this assumption is explained in the next subsection. The validity of this hypothesis is verified in Section~\ref{sec: validation} where independent comparisons with experimental data are presented. 

\subsection{The energy storage based on crosslink density evolution} \label{subsec: energy storage}
Assuming that the specific internal energy and entropy of the oxygen inside the solid medium are negligible, the second law of thermodynamics in the form of the Clausius–Duham inequality for a solid medium can be presented as
\begin{align} \label{eq:CD}
\frac{1}{2}\mathbf{S} \cdot \dot{{\mathbf{C}}}  - \rho \dot \Psi - \rho \dot T Z  -  \frac{\mathbf{Q} \cdot \nabla T}{T} \geq 0
\end{align}
where ${\mathbf{C}}$ is the right Cauchy-Green strain tensor, $\mathbf{S}$ is the second Piola-Kirchhoff stress tensor, $\rho$ is the density, $\Psi$ is the specific Helmholtz free energy, $Z$ is the the specific entropy, $T$ is temperature, and $\mathbf{Q}$ is Lagrangian heat flux. If we assume that the Helmholtz free energy is a function of the right Cauchy-Green strain tensor and the temperature of the medium, i.e., $\Psi = \Psi \left( { {\mathbf{C}}}, T \right) $, and using the chain rule to take the derivative of the Helmholtz free energy and substitute it in Eq. (\ref{eq:CD}), the Clausius–Duham inequality becomes
\begin{align} \label{eq: Helmholtz in CD-0}
\left(\frac{1}{2}\mathbf{S} - \rho \frac{\partial \Psi}{\partial \mathbf{C} }  \right)  \cdot \dot{\mathbf{C}}  -  \rho \left(  \frac{\partial \Psi}{\partial T} + Z \right) \dot T  -  \frac{\mathbf{Q} \cdot \nabla T}{T} \geq 0
\end{align}
It is known that since the positivity of energy dissipation must hold for all processes, we can get
\begin{align} \label{eq: Helmholtz in CD}
\mathbf{S} = 2\rho \frac{\partial \Psi}{\partial  {\mathbf{C}} }, \quad \text{and} \quad Z= -\frac{\partial \Psi}{\partial T}  
\end{align}

Therefore, the constitutive equation can be obtained by a correct assumption of the Helmholtz free energy, or how the elastomer stores energy upon the thermo-chemical aging phenomena. In the general and robust two-potential point framework, the forms that material stores and dissipates energy must be assumed or obtained in order to develop an accurate constitutive equation. It should be mentioned again that in this work, we assume that the energy dissipation in the thermo-chemical aging process can be neglected. We assume that the dominant thermo-chemical aging phenomenon is the extra crosslinks formation within the elastomer, which is an energy storage process. We modify the network stiffness and the chain extensibility in the well-known Arruda-Boyce eight-chain model \citep{ArrudaBoyce93} to incorporate the additional energy stored within elastomers due to thermo-chemical aging processes. The Helmholtz free energy associated with oxygen diffusion is neglected here as well. It must be noted that in most of the previous studies where the chemical Helmholtz free energy was included, the term was only used to obtain the diffusion equation. A full coupling between the oxygen diffusion and mechanical loading is basically rare and is justified by the different time frames of diffusion and mechanical loading, as the diffusion process is very slow. 

The Arruda-Boyce eight-chain constitutive equation accounts for the non-Gaussian nature of the molecular chain stretch and provides an accurate representation of the large-strain behavior of rubber-like materials under different states of deformation. An attractive feature of the eight-chain Arruda-Boyce model (besides being micro-mechanically motivated) is that it only requires two physics-based material properties, i.e., the network chain density (or equivalently the rubber shear modulus), and the limiting chain extensibility to model elastomer behavior under different states of deformation (i.e., uniaxial, shear, and biaxial). The Helmholtz free energy according to the incompressible Arruda–Boyce model is given by \citep{ArrudaBoyce93}
\begin{align} \label{eq: Helmholtz function for Arruda full form} 
\Psi_{AB} \left(\mathbf{C}\right) = n_0 K_B \Theta N_0\Bigg[ \frac{\lambda_{chain}}{\sqrt{N_0}} \mathcal{L}^{-1}\Big(\frac{\lambda_{chain}}{\sqrt{N_0}}\Big) + ln\frac{\mathcal{L}^{-1}\Big(\frac{\lambda_{chain}}{\sqrt{N_0}}\Big)}{sinh(\mathcal{L}^{-1}\Big(\frac{\lambda_{chain}}{\sqrt{N_0}}\Big))} \Bigg]
\end{align}
where $n_0$, $K_B$, and $\Theta$ are the initial number of chains per unit volume, the Boltzmann constant, and the absolute temperature, $N_0$ is the number of Kuhn monomers per chain of the elastomer and is related to the limiting chain extensibility $\lambda_{lock}$ as $N_0 = \lambda_{lock}^2 $, $\mathcal{L}(\cdot) = coth(\cdot) - \frac{1}{(\cdot)}$ is the Langevin function whose inverse $\mathcal{L}^{-1}$ is given by several approximations in the literature, among which we find the well-known Pade approximation that is $\mathcal{L}^{-1}(x) = x \frac{3-x^2}{1-x^2}$, and $\lambda_{chain}(\mathbf{C}) = \sqrt{\frac{I_1(\mathbf{C})}{3}}$ is the relative macro-stretch written as a function of the first invariant of the right Cauchy-Green strain tensor $I_1(\mathbf{C})=tr(\mathbf{F}^T\mathbf{F})$, where $\mathbf{F}$ is the deformation gradient tensor. Note that the dependence of the strain energy expression for the Arruda-Boyce eight-chain constitutive equation on the first invariant of the right Cauchy-Green tensor makes it an invariant based hyperelastic constitutive equation.

Equation~\ref{eq: Helmholtz function for Arruda full form} can be written in polynomial form using the first five terms of the inverse Langevin function as
\begin{align} \label{eq: Helmholtz function for Arruda} 
\Psi_{AB} \left(\mathbf{C}\right) = \mu_0 \mathlarger{\sum}_{i=1}^5 c_i  \frac{1}{N_0^{2i-2}} \left( I_{1\mathbf{C}}^i -3^i  \right)
\end{align}
where $\mu_0 = n_{0} K_B \Theta$ is the rubber shear modulus.  
The constants $c_i$ in Equation~\ref{eq: Helmholtz function for Arruda} are equal to $  c_1=\tfrac{1}{2}, c_2= \tfrac{1}{20}, c_3= \tfrac{11}{1050}, c_4= \tfrac{19}{7000}, c_5= \tfrac{519}{673750}$.

As it has been argued in subsection~\ref{subsec: thermo-chemical aging mechanism}, crosslink formation is the dominant mechanism in the thermo-chemical aging of elastomers, and an oxidized product has a more crosslinked network of smaller chains compared to the unaged polymer with the longer crosslinked network \citep{KONICA2021104347}. Therefore, the number of Kuhn monomers per chain in elastomers changes and decreases upon aging. On the other hand, the formation of crosslinks between these newly formed short-chains induces more stiffness as the deformation of short polymer chains in a highly crosslinked material is more difficult. Recognizing that the Arruda-Boyce Helmholtz free energy is indeed a function of two main micro-mechanically motivated material parameters: (1) the rubber shear modulus $\mu_0$ (which increases with respect to aging time) and (2) the number of Kuhn monomers per chain $N_0$ (which decreases with respect to aging time), the effect of aging on the Helmholtz free energy can be accounted for by describing appropriate evolution functions of $\mu$ and $N$ with respect to aging time, i.e., $\mu(t)$ and $N(t)$.

First, motivated by the expression of the shear modulus in Equation~\ref{eq: Helmholtz function for Arruda}, $\mu_0 = n_{0} K_B \Theta$, we consider that the increase in the number of the newly formed crosslinks per volume due to aging (i.e., crosslink density) directly affects the shear modulus of the material at the corresponding aging state and write a micro-mechanically motivated expression for the evolution of $\mu(t)$ as
\begin{align} \label{eq: evolution of mu - eq1} 
\begin{split}
 \mu(t) & = n_{0} K_B \Theta + \big(\nu(t) - \nu_0\big)R \Theta \\
        & = \mu_0 + \big(\nu(t) - \nu_0\big)R \Theta
\end{split}
\end{align}
where $\nu_0$ and $\nu(t)$ are the crosslink densities of the unaged material (at aging time t=0) and the aged material (at some later aging time t), respectively, and $R$ is the natural gas constant. In equation~\ref{eq: evolution of mu - eq1}, the term $(\nu - \nu_0)$ gives the change in the crosslink density between the primary network configuration and the network configuration corresponding to a given aging state. A stiffness-like component is introduced by multiplying the change in the crosslink density which has units of moles per volume by the natural gas constant $R$ and the absolute temperature $\Theta$. 

Second, the total number of crosslinks per volume times the number of Kuhn segments per chain must remain constant in order to satisfy the conservation of mass principle. 
Assuming that individual chains do not diffuse within the elastomer \citep{holzapfel2000nonlinear} and that the elastomer is incompressible, 
the following equilibrium holds
\begin{align} \label{eq: Chain conservation} 
N(t) \nu(t) = N_0 \nu_0
\end{align}
where $N(t)$ is the number of Kuhn monomers per chain of the elastomer at the current state of aging. As a result of these two modifications, the additional energy storage due to crosslink density evolution is effectively considered and the expected stiffness variation upon aging is appropriately accounted for.

Following the macromolecular network evolution motivated expressions for $\mu(t)$ and $N(t)$, we can now write the following physics-based constitutive equation to incorporate the additional stored energy inside the material due to crosslink formation induced by thermo-chemical aging as
\begin{align} \label{eq: assumed Helmholtz aging}
\Psi_{a} \left(\mathbf{C}\right) = \mu(t) \mathlarger{\sum}_{i=1}^5 c_i  \frac{1}{N(t)^{2i-2}} \left( I_{1\mathbf{C}}^i -3^i  \right)
\end{align}
where $\mu(t)$ and $N(t)$ are given by Equations~\ref{eq: evolution of mu - eq1} and~\ref{eq: Chain conservation}, respectively, and $\Psi_{a}$ is the Helmholtz free energy of the aged material combining the contribution from the original elastomer network prior to aging and the contribution of the secondary network formed due to crosslinking. Substituting Equation~\ref{eq: assumed Helmholtz aging} into Equation~\ref{eq: Helmholtz in CD}a gives the second Piola-Kirchhoff stress tensor. The first Piola-Kirchhoff stress tensor $\textbf{P}$ can then be computed as $\mathbf{P} = \mathbf{F} \mathbf{S}$.

%

A similar approach in considering the evolution of the network concentration with respect to aging had been introduced in the literature. Several researchers assumed a temporal evolution of the material properties to incorporate the variation in stiffness upon subjecting the elastomer to different multi-physics phenomena like curing or thermo-oxidative aging (\cite{dal2009micro,hossain2015continuum}). For instance, \citet{hossain2015continuum} used an evolution function, including three model parameters, for the shear modulus to describe the exponential increase in stiffness due to curing processes. However, defining characteristics that make our model more distinguishable are that, first, the evolution of the shear modulus in our constitutive equation depends directly on the crosslink density achieved at a given aging state, and second, the changes in the number of Kuhn monomers per chain is also considered. Therefore, the developed constitutive equation herein provides a one-to-one mapping between chemically-based quantities and physically-based macroscopic variables. The developed constitutive equation relies on two material properties (i.e., $\mu_0$ and $N_0$) in addition to the crosslink density evolution data. In the case of using an equation to consider the crosslink density evolution, an s-shape function that requires three model parameters can be used.


The obtained constitutive equation follows the consistency conditions as: (1) $\Psi_a \left(\mathbf{C}\right) =0$ when $\mathbf{C}=\mathbf{I}$, that is, when a network is in its reference configuration; (2) $\Psi_a \left(\mathbf{C}\right) > 0 $ when $\mathbf{C} \neq \mathbf{I}$, that is, when a network is deformed; (3) each network is stress free in its reference configuration and the network reformation process occurs stressless. It should also be mentioned that the original network in the eight-chain model was assumed to be isotropic in its reference configuration. Subsequently, each newly formed crosslink is also assumed to be isotropic in its reference configuration. The assumption of isotropic crosslink formation and network evolution is only made because of the absence of experimental results about the symmetry of new networks. This simplifying assumption should be modified to account for a different material symmetry once new understandings of the network evolution may be found. It also should be mentioned that aging during the curing process \citep{MAHNKEN20132003}, and possible crack healing in elastomers \citep{kumar2018fracture} are out of the scope of this study. 

A summary of the simplifying assumptions in this work is: (1) the energy storage and crosslinking mechanism is the dominant phenomenon in thermo-chemical aging of elastomers, (2) the elastomer is saturated with oxygen and the energy due to chemical diffusion can be neglected, (3) the chemical reactions occur homogeneously inside the thin samples, (4) temperature does not change locally upon the application of mechanical loading, or is negligible (5) the material is assumed incompressible and isotropic, and (6) the aging process does not depend on the strain. Most of these assumptions are common among all the constitutive equations which were developed to simulate thermo-chemical aging in elastomers in the literature (e.g., \cite{SHAW20052758,Johlitz2013,MOHAMMADI2020109108}). However, the developed constitutive equation in this work can be easily incorporated into the more complicated diffusion-reaction-based constitutive equations to be fully coupled with the diffusion equations. The future work of the authors is twofold. First, to incorporate the obtained Helmholtz free energy into a diffusion-reaction and thermodynamic-based framework to simulate a fully coupled thermo-chemo-mechanical aging. Second, to incorporate the effect of chain-scission and energy dissipation for a more accurate representation of aging phenomena based on other chemical characterization tests such as nuclear magnetic resonance (NMR) and Fourier-transform infrared spectroscopy (FTIR).  

\section{Validation of the developed thermo-chemical aging constitutive equation} \label{sec: validation} \label{cont model validation}
In this section, the prediction of the proposed constitutive equation is validated against a set of experimental data available in the literature to assess its capabilities \citep{hamed1999tensile,ZHI201915,Abdelaziz2021}. This particular set of literature was chosen such that the reported articles contained information about both the crosslink density evolution upon aging and a mechanical response of aged samples. Two types of elastomer compounds (i.e., SBR and NR) with various Carbon Black filler fractions -- labeled unfilled and filled -- were identified. Samples were subjected to thermo-chemical aging at various temperatures ranging from $70^{\circ}C$ to $120^{\circ}C$.

First, \citet{hamed1999tensile} reported the evolution of crosslink density and tensile tests for filled and unfilled SBR and NR. The specimens were subjected to air-oven aging at 100$^{\circ}$C for various aging times. All specimens were prepared with an average thickness below the critical thickness for diffusion-limited oxidation (thicknesses varied between 0.15 and 0.2~$mm$) to ensure uniform oxidation. The specimens were subsequently tested in the tensile mode under a strain rate of approximately 0.017$s^{-1}$. The crosslink densities for the unfilled materials were calculated using the classic Flory-Rehner equation, whereas, for the filled samples, the Kraus modification was employed. In the second study, \citet{ZHI201915} reported crosslink density evolution as well as tensile test results of a filled SBR material for several aging times. In their study, SBR samples were thermally aged in an air-circulated oven at 120$^{\circ}$C. The monotonic stretching tests on samples with different aging times were conducted at a stretch rate of 0.06$s^{-1}$. 
In the last study, \citet{Abdelaziz2021} reported crosslink density evolution for a filled SBR material aged under an accelerated aging process at different temperatures and for different exposure times. Tensile mechanical test results were conducted by a stretch rate of 0.0055$s^{-1}$ and reported for all aging times.

In the following for each data set, the two mechanical material parameters, $\mu_0$ and $N_0$, were first obtained based on the unaged tensile stress experimental data and the one-dimensional uniaxial Arruda-Boyce constitutive equation. Table~\ref{Table: unaged model parameters} presents the values for $\mu_0$ and $N_0$ corresponding to each particular unaged elastomer compound. Then, the evolution of the crosslink density obtained experimentally for various elastomers was used to predict the stress-strain responses based on Equations~\ref{eq: evolution of mu - eq1}--\ref{eq: assumed Helmholtz aging} and the first Piola-Kirchhoff stress definition. The average prediction error for each aging duration and each elastomer compound was calculated according to
\begin{align} \label{eq: error}
\text{Average error} = \text{average} \left( \frac{\text{experimental data} - \text{model prediction}}{\text{experimental data}} \right) \times 100 
\end{align}
where the experimental data and model prediction are data points calculated at select strain levels in the stress-strain response.

\begin{table}[h]
\centering
\caption{The shear modulus and the number of Kuhn monomers per chain for the unaged elastomers obtained by fitting the Arruda-Boyce eight-chain constitutive equation to the unaged stress-strain curves.}
\resizebox{\columnwidth}{!}{%
\begin{tabular}{lccccc} 
        \toprule
        & \multicolumn{4}{c}{\citet{hamed1999tensile} ($T=100^{\circ}C$)} \\
        {} & Unfilled SBR & Filled SBR & Unfilled NR & Filled NR \\
        \cmidrule{2-5}
        $\mu_0~(MPa)$ & 0.58 & 1.79 & 0.5 & 2 \\
        $N_0$ & 140 & 2.961 & 11.5 & 4.5 \\ \midrule
        & \citet{ZHI201915} & \multicolumn{3}{c}{\citet{Abdelaziz2021}} \\
        {} & $T=120^{\circ}C$ & $T=100^{\circ}C$ & $T=90^{\circ}C$ & $T=70^{\circ}C$ \\
        \cmidrule{2-5}
        $\mu_0~(MPa)$ & 0.64 & 0.58 & 1.79 & 0.5 \\
        $N_0$ & 1.5 & 140 & 2.961 & 11.5 \\ 
     \cmidrule{1-5}
    \end{tabular}
\label{Table: unaged model parameters}
}
\end{table}


\subsection{SBR} \label{subsec: SBR validation}
\begin{figure*}[h!bt]
    \centering
        \subfloat[]{%
        \includegraphics[width=0.48\linewidth]{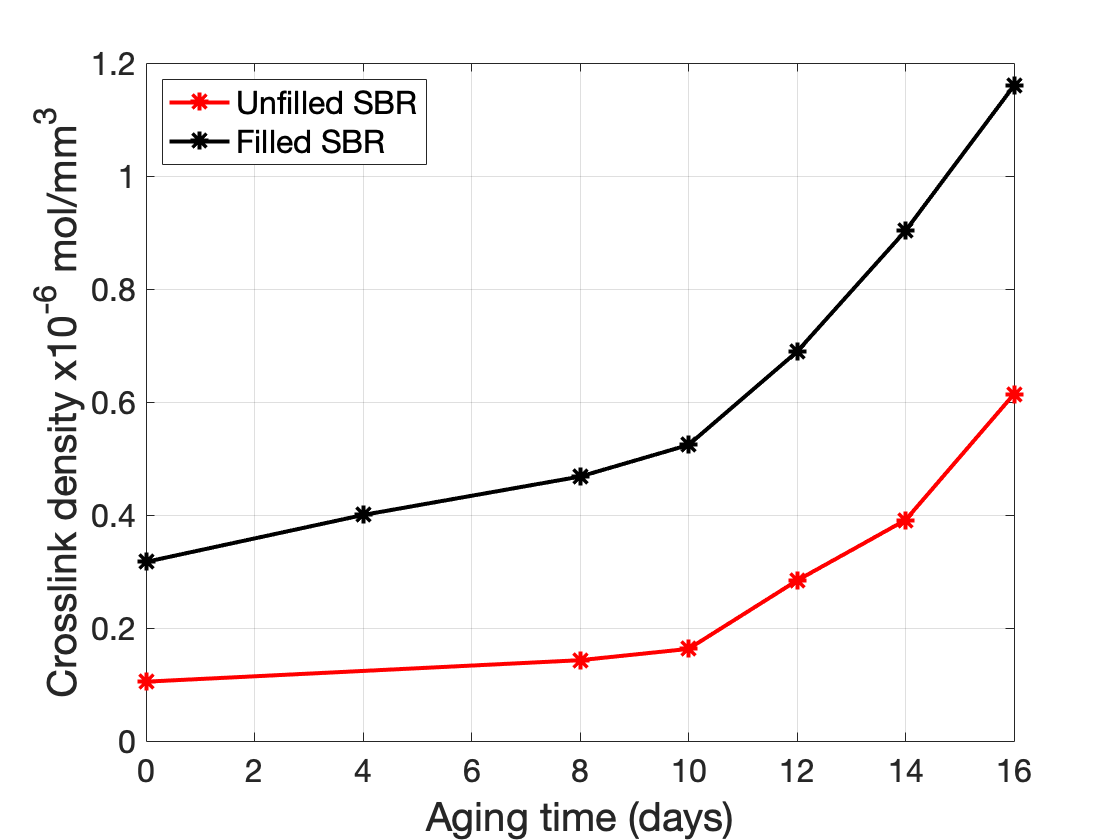}
        \label{fig1:a}
        }%
    \hfill 
    \subfloat[]{%
        \includegraphics[width=0.48\linewidth]{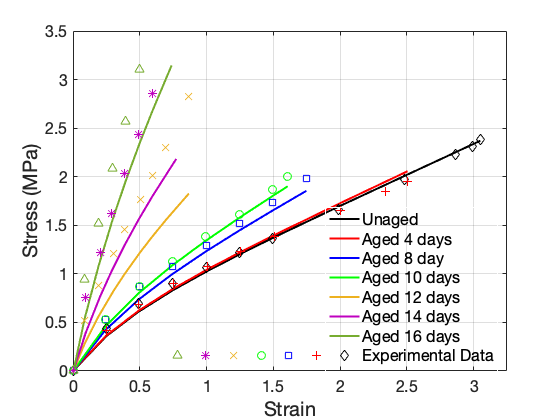}
        \label{fig1:b}
        }%
    \hfill
    \subfloat[]{%
        \includegraphics[width=0.48\linewidth]{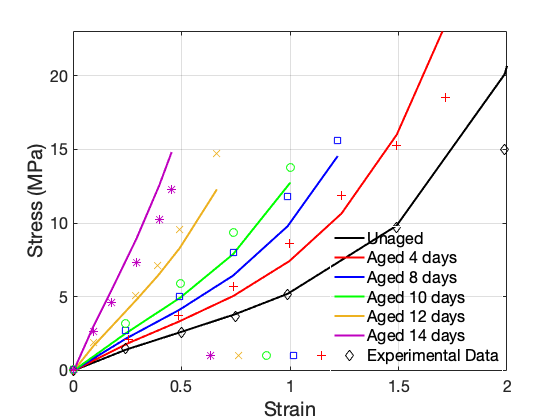}
        \label{fig1:c}
        }%
    \caption {a) Crosslink density evolution as the aging time, and predictions of the developed constitutive equation for different aging times of b) unfilled SBR and c) filled SBR. (The experimental data were reproduced based on \citet{hamed1999tensile}.)}
    \label{fig1: Hamed SBR}
\end{figure*}

\begin{figure*}[h!bt]
    \centering
    \subfloat[]{%
        \includegraphics[width=0.48\linewidth]{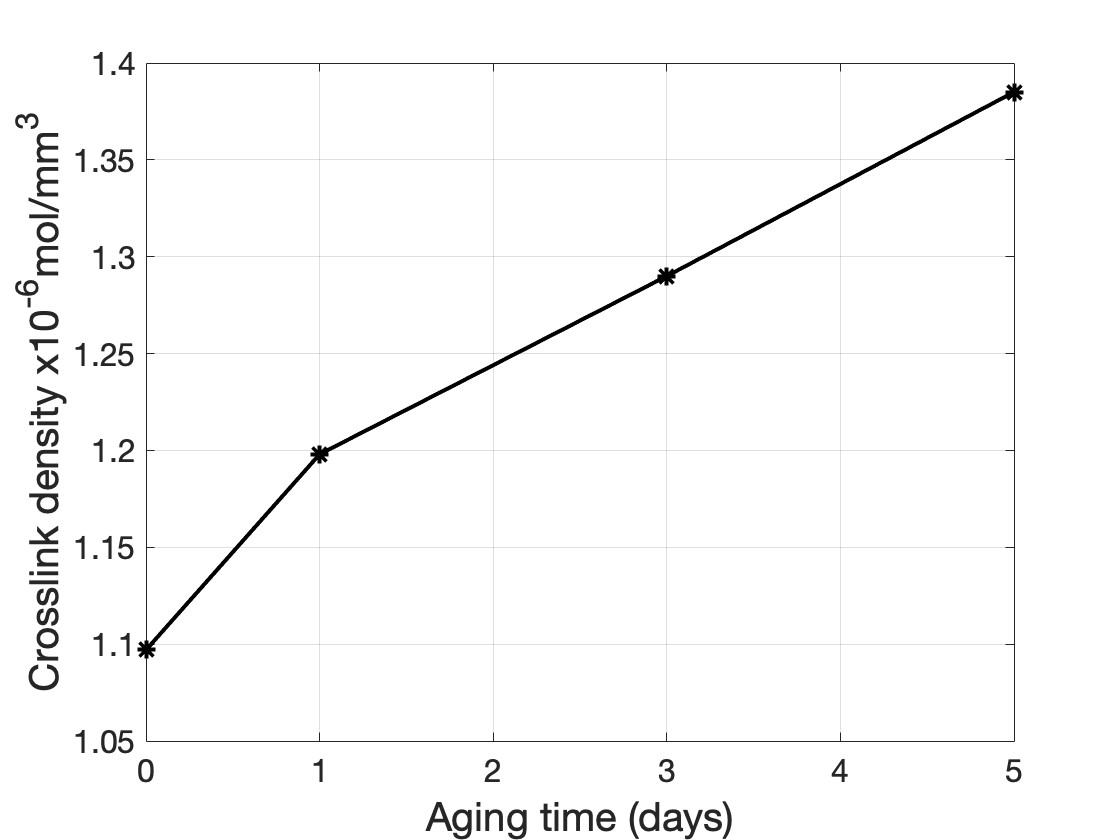}
        \label{fig2:a}
        }%
    \hfill
    \subfloat[]{%
        \includegraphics[width=0.48\linewidth]{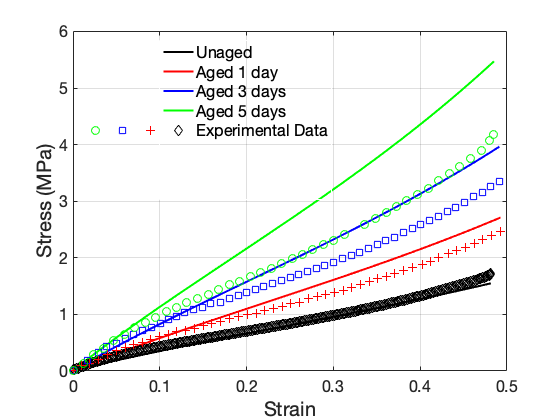}
        \label{fig2:b}
        }%
    \caption {a) Crosslink density evolution as the aging time and b) predictions of the developed constitutive equation. (The experimental data were reproduced based on \citet{ZHI201915}.) } 
    \label{fig: Zhi SBR}
\end{figure*}

\begin{figure*}[h]
    \centering
    \subfloat[]{%
        \includegraphics[width=0.48\linewidth]{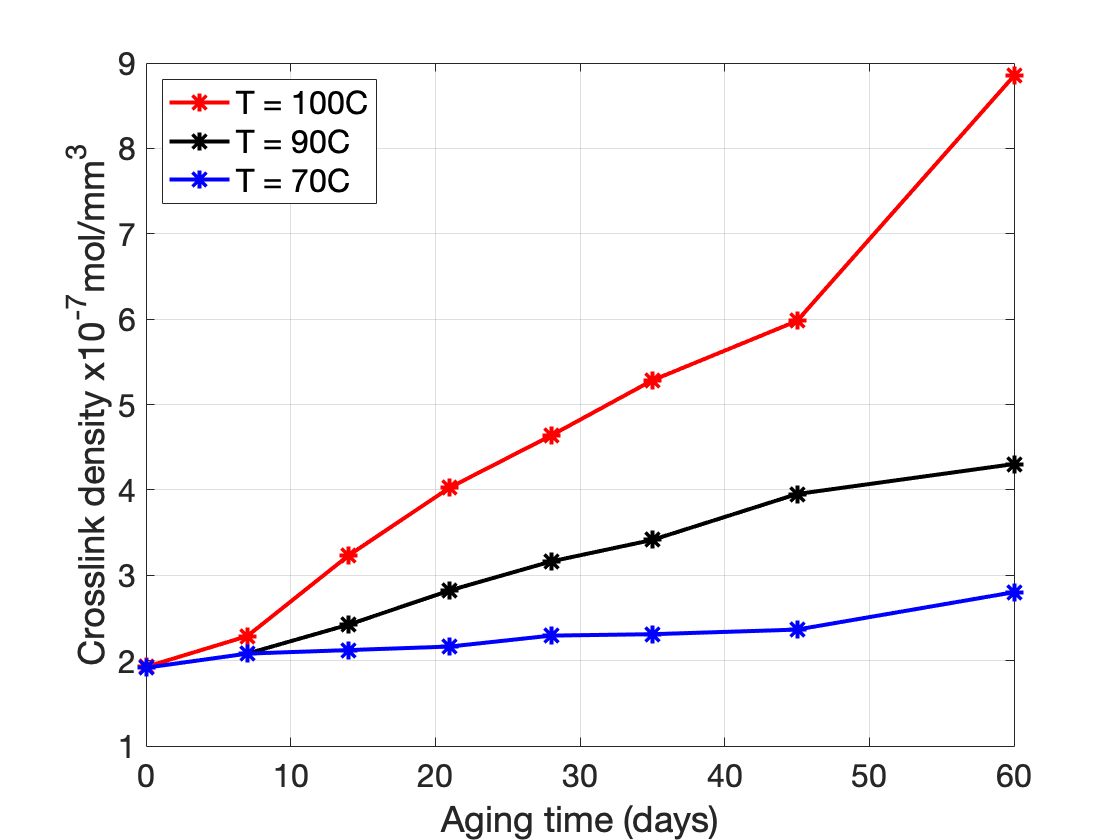}
        \label{fig3:a}
        }%
    \hfill
    \subfloat[]{%
        \includegraphics[width=0.48\linewidth]{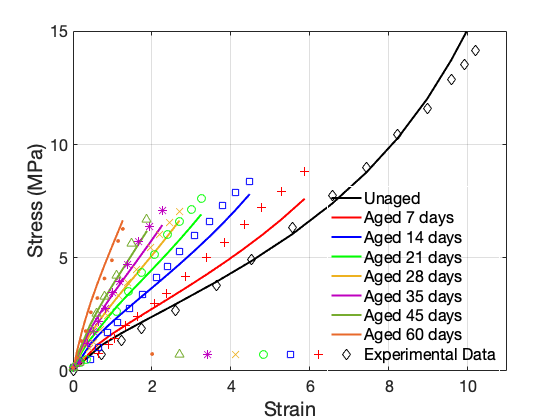}
        \label{fig3:b}
        }%
    \hfill
    \subfloat[]{%
        \includegraphics[width=0.48\linewidth]{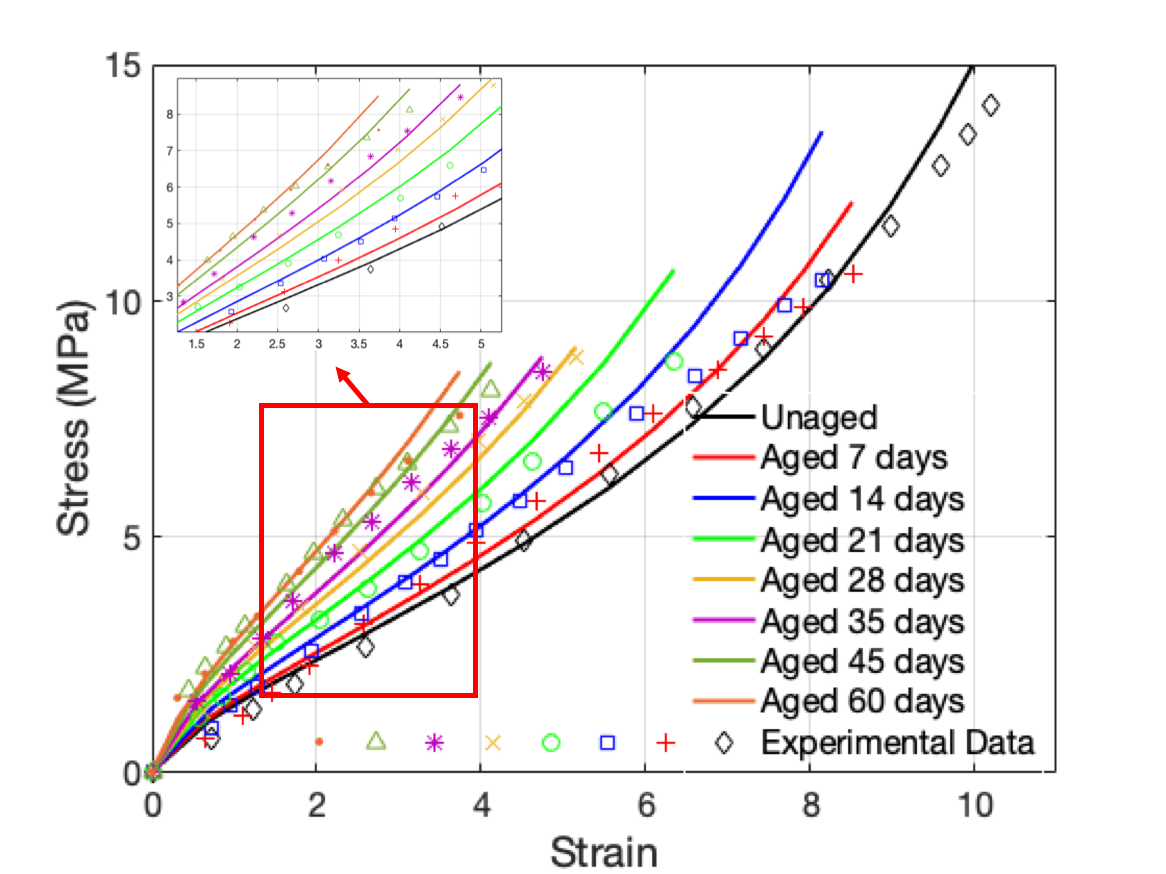}
        \label{fig3:c}
        }%
      \hfill
    \subfloat[]{%
        \includegraphics[width=0.48\linewidth]{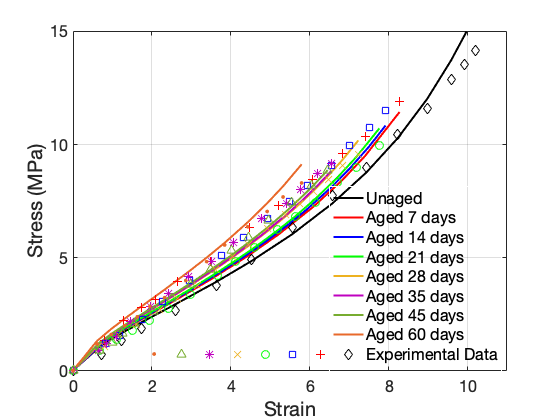}
        \label{fig3:d}
        }%
    \caption {a) Crosslink density evolution as the aging time at different temperatures for a filled SBR elastomer; and predictions of the developed constitutive equation for different aging time for the cases when the aging temperature is equal to b) $100^{\circ}$C, c) $90^{\circ}$C, and d) $70^{\circ}$C. (The experimental data were reproduced based on  \citet{Abdelaziz2021}.)} 
    \label{fig: Abdelaziz SBR}
\end{figure*}

In this section, the available data in the literature are used to predict the uniaxial tensile stress-strain responses of unfilled and filled SBR based on their crosslink density evolution upon aging. Figures~\ref{fig1: Hamed SBR},~\ref{fig: Zhi SBR}, and~\ref{fig: Abdelaziz SBR} demonstrate the model prediction for SBR materials corresponding to the studies of \citet{hamed1999tensile}, \cite{ZHI201915}, and \citet{Abdelaziz2021}, respectively. For each study and its corresponding figure, both the crosslink density evolution as well as the model predictions are presented. Moreover, the average calculated errors using Equation~\ref{eq: error} are also summarized in Table~\ref{Table: errors}. 

It can be observed in Figure~\ref{fig1: Hamed SBR} that the evolution of the crosslink density follows the same pattern as the change in stiffness for each given aging state wherein the modulus evolves in accordance with the change in the crosslink density. The modulus -- or equivalently stiffness --  increases substantially after an initial aging time during, mimicking the behavior observed for the crosslink density. The parallelism in the evolution of the elastomer's crosslink density and its corresponding stiffness reinforces the argument  that crosslinking reactions are more dominant during thermo-chemical aging, especially at larger aging times. 


Moreover, regarding the \cite{hamed1999tensile} unfilled SBR data, it can be seen that errors remain below $10\%$ for the lower aging times (i.e., up to 10-day aging). As the aging time increases, the response for the unfilled SBR predicted by the proposed constitutive equation underestimates the experimental observations. The underestimation could likely be attributed to additional molecular chain segments that have potentially been created between free radicals and double C-C bonds in the unfilled SBR. The additional chains were not accounted for by the current developed constitutive equations. These additional chain segments possess the same load-resisting capacity as the original network chains (albeit shorter in length) and contribute to an increase in the chain density in the unfilled SBR. Indeed, researchers have argued that two types of reactions can participate during thermo-chemical aging processes: reaction of radicals with double C-C bonds and reaction of radicals among themselves \citep{mohammadi2019micro,le2016predictive}. From what it appears, the proposed constitutive relationship accounts for the latter type and therefore results in underestimating the material stiffness, especially at higher aging times.


On the other hand, the predicted response for the case of \citet{hamed1999tensile}'s filled SBR -- which contains 50 phr of carbon black -- is quite accurate even at higher aging times. The inclusion of carbon black inhibits radical reactions with double C-C bonds and thus reduces the potential of creating additional chain segments between the corresponding junction ends. At the same time, however, carbon black increases the probability of recombination of primary macro-radicals and promotes the creation of additional crosslinks in the material. This is consistent with arguments provided in the literature, which suggest that carbon black acts as a free-radical scavenger capable of binding macro-radicals generated during oxidation \citep{janowska2008influence,hamed1999tensile}. Those additional load-resisting chain segments manifested for the unfilled SBR are therefore absent for the filled SBR. The increase in the density of crosslinks at radical-radical junctions is the major contributor to the stiffening behavior of filled SBR, as is properly accounted for by the proposed constitutive relationship. 

The effects of carbon black on the predictions are consistent with the results of \citet{Abdelaziz2021} shown in Figure~\ref{fig: Abdelaziz SBR}. The SBR utilized in their study contained 34 phr of carbon black, and therefore, represents an intermediate value between those corresponding to the SBR materials used in \citet{hamed1999tensile} (i.e., 0 phr for unfilled and 50 phr for filled). At the similar aging temperature of $100^{\circ}C$, the average errors predicted for this intermediate carbon black fraction fall between those obtained for the cases where the carbon black was 0 and 50 phr. In other words, the average error decreases with an increase in the carbon black volume fraction. Thus, reinforcing the argument that at higher carbon black contents, the potential of additional load-resisting chains forming due to aging is reduced, and the number of crosslinks per elastomer volume is increased. 




Figure~\ref{fig: Abdelaziz SBR} illustrates the predictions of the \citet{Abdelaziz2021} data at three different aging temperatures of $70^{\circ}C$, $90^{\circ}C$, and $100^{\circ}C$. It can be observed that the aging temperature also affects the prediction capability of the proposed constitutive equation. The proposed constitutive equation can replicate the filled SBR's experimental responses very well for the lower aging temperatures (i.e., $70^{\circ}C$, $90^{\circ}C$). In fact, the average error remains below 13\% and 14\% when the aging temperature is $70^{\circ}C$ and $90^{\circ}C$, respectively. As the aging temperature increases, the constitutive equation tends to overestimate the response of filled SBR relatively compared to the lower temperatures. The overestimation can be explained by the fact that the current model neglects temperature-dependent dissipation mechanisms such as viscoelasticity and other degradation events such as extended chain-scission and crosslink breakage, which likely become more considerable at higher aging temperatures. The overestimation of the experiments can also be observed in the case of \citet{ZHI201915} data as well, where the specimens were aged at $120^{\circ}C$. Thus, although the SBR used in \citet{ZHI201915}'s study contains the same carbon black volume fraction as the one used in \citet{hamed1999tensile}, the difference in the calculated average errors between the two studies is attributed to the increased aging temperature (i.e., $100^{\circ}C$ compared to $120^{\circ}C$). Nevertheless, it is worth mentioning that the experimental stress-strain responses corresponding to the work of \citet{ZHI201915} focused on the lower stress-strain intervals as opposed to the other works where the authors presented the complete stress-strain behavior. The lack of the complete nominal stress-strain behavior could potentially have affected our prediction results. 

Based on the outcome, it can be concluded that the proposed constitutive relationship is most adequate for predicting the stiffness of filled SBR as the potential for forming new macromolecular chains is reduced for these carbon black-filled materials compared to their unfilled counterparts. However, at higher aging temperatures (i.e., temperature greater than 100$^{\circ}C$), the effects of temperature-dependent dissipation mechanisms such as viscoelasticity or other molecular dissipation events such as extended chain-scission events become more dominant, and the prediction using the proposed model is relatively overestimated. 

\subsection{NR} \label{subsec: NR validation}
\begin{figure*}[h!bt]
    \centering
        \subfloat[]{%
        \includegraphics[width=0.48\linewidth]{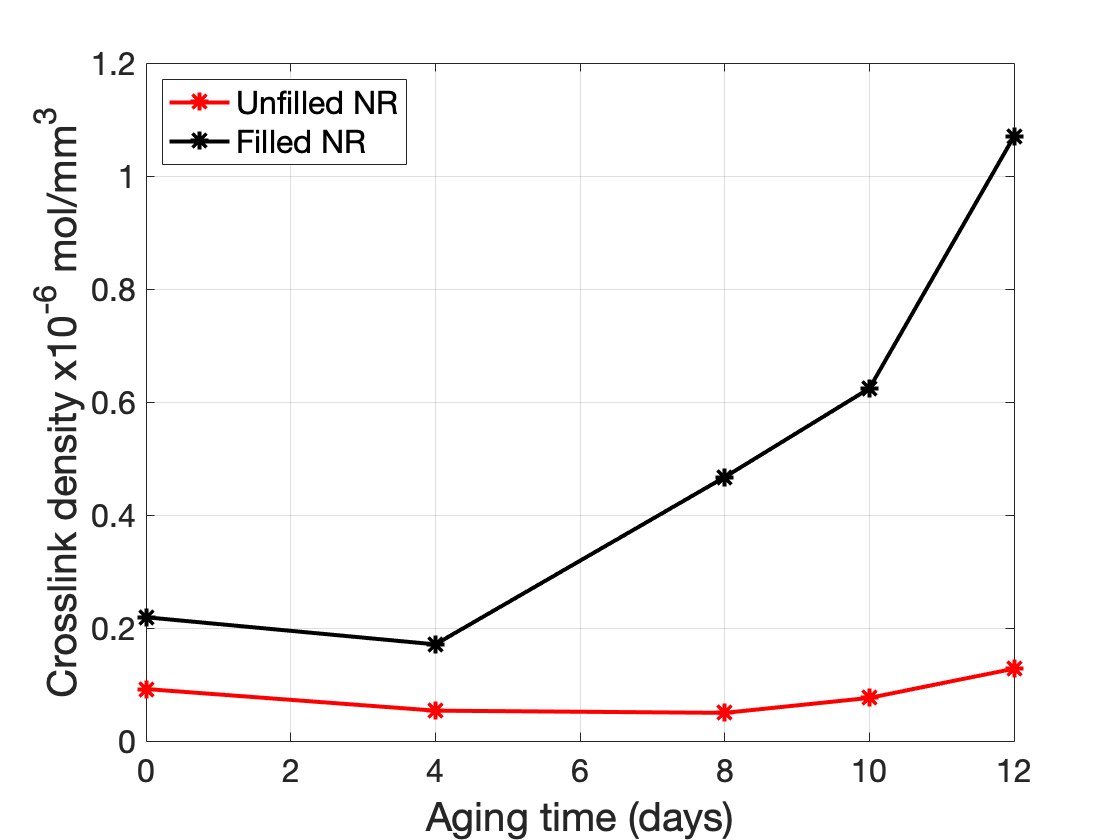}
        \label{fig4:a}
        }%
    \hfill 
    \subfloat[]{%
        \includegraphics[width=0.48\linewidth]{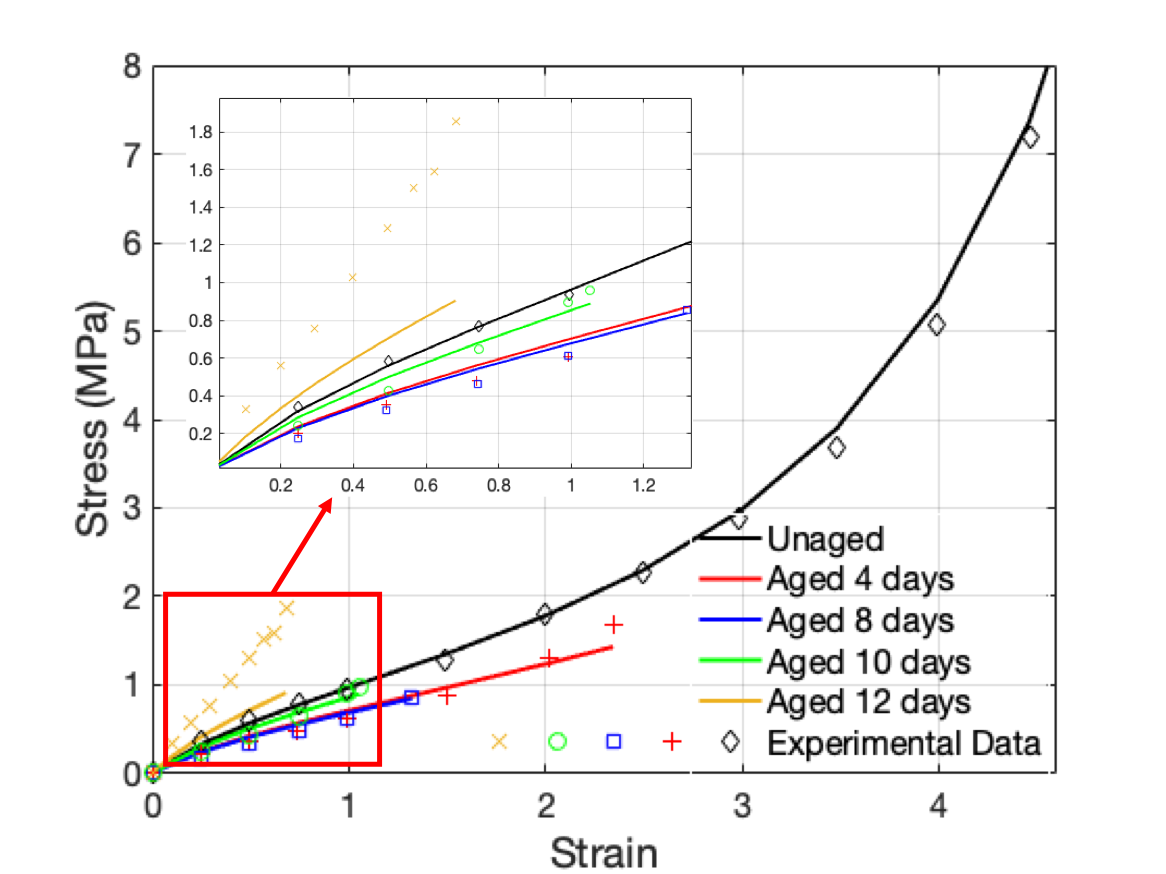}
        \label{fig4:b}
        }%
    \hfill
    \subfloat[]{%
        \includegraphics[width=0.48\linewidth]{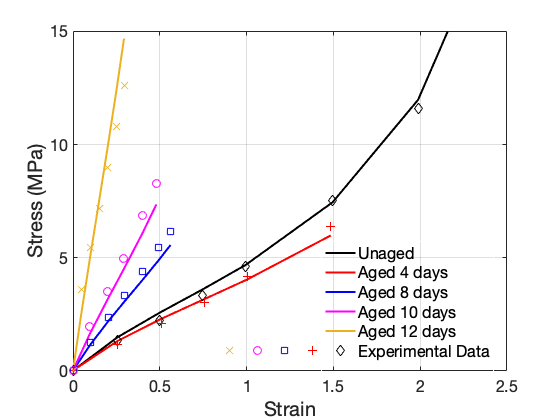}
        \label{fig4:c}
        }%
    \caption {a) Crosslink density evolution as the aging time, and predictions of developed constitutive equation for different aging time of b) unfilled NR and c) filled NR. (The experimental data were reproduced based on \citet{hamed1999tensile}.)}
    \label{fig: Hamed NR}
\end{figure*}

In this section, the capability of the proposed constitutive equation in predicting the tensile stress-strain responses of unfilled and filled NR at various aging states \citep{hamed1999tensile} is outlined. Figure~\ref{fig: Hamed NR} demonstrates the evolution of the crosslink density for both unfilled and filled NR as well as the corresponding model predictions compared to the experimental results. It can be seen that the developed constitutive equation can predict the responses very well regardless of whether the material stiffens or softens. As evidenced by Figure~\ref{fig: Hamed NR}b, unfilled NR undergoes an extended period of softening up to 10 days. During this aging stage, the crosslink density diminishes compared to the initial unaged state. The dependence of the stiffness parameter on the crosslink density makes it possible for the constitutive equation to capture the case when the material actually softens (i.e., the case where network chain disruption events are greater than its formation counterparts). Additionally, similar to the behavior observed regarding the effect of carbon black on SBR's predictions, the predicted responses for the filled NR are more accurate compared to the unfilled NR at higher aging duration. In fact, the highest calculated error for filled NR was approximately 13.9\% for the 12-day aging time (compared to about 44\% for the unfilled NR at the same aging time). 


\subsection{Discussion}
The proposed constitutive equation can, given the evolution of the crosslink density, predict the mechanical responses of aged elastomeric materials fairly well even when only two material parameters are employed -- in this case, the rubber shear modulus and the number of Kuhn monomers per chain of the unaged elastomer within the Arruda-Boyce eight-chain constitutive theory. The ability of the constitutive equation to accurately predict the mechanical test results of aged materials independent of any mechanical tests, i.e., without conducting any further fitting to the experimental test results on aged samples, constitutes an attractive feature of the proposed relationship. The average reported errors in this study with the low number of model parameters is unprecedented in the literature. Many, if not all of the existing works in the literature, use several mechanical tests to fit and obtain numerous model parameters associated with their constitutive equations. 

For example, \cite{JOHLITZ2014138}'s phenomenological constitutive equation is based on the Mooney-Rivlin hyperelastic theory and contains fourteen model parameters. Their constitutive model, which is based on the concept of state variables considering network scission and reformation events, requires two sets of mechanical experimental tests for different aging conditions for calibration. 
Moreover, recently, \citet{MOHAMMADI2020109108} used three to six parameters for an Arrhenius-based aging decay function. In addition to the Arrhenius-based aging decay function parameters, seven mechanical model parameters were used to fit the experimental observations. Their average errors at T=95$^{\circ}$C, which were deemed acceptable, were computed to be 5.75\%, 21.55\%, 23.52\%, and 12.01\% for the unaged, one-day, five-day, and 10-day aging times, respectively. Since our average error range falls below these values, it can be argued that the developed constitutive equation herein yields acceptable predictions even when very few material properties and tests are utilized. Moreover, most kinetics models available in the literature employ a series of model parameters to describe only one mechanism at a time (e.g., either when the aged material is softer than its unaged configuration or vice-versa). Doing so results in a large number of variables to appropriately consider all cases of network evolution, such as in the case of \cite{LION20121227}. The crosslink density evolution approach adopted in this work bypasses the need for incorporating more material parameters than necessary and yields reasonable estimations of the response of aged elastomers, especially for the case of carbon black-filled elastomers. The proposed constitutive equation avoids the need for conducting further mechanical tests and provides predictions of the responses of aged materials within reasonable accuracy using only two material parameters corresponding to the unaged configuration in addition to one chemical characterization test.


\begin{table}[h!bt]
\centering
\caption{Associated average errors obtained by comparing the predicted stress-strain results based on Equations~\ref{eq: evolution of mu - eq1}-\ref{eq: assumed Helmholtz aging} and experimental tensile test results.}
\begin{tabular}{lccccccc} 
        \toprule
        \multicolumn{8}{c}{\citet{hamed1999tensile} SBR} \\
        {} & Unaged & 4-day & 8-day & 10-day & 12-day & 14-day & 16-day \\
        \midrule
        Unfilled & 4.68 & 5.42 & 8.94 & 5.33 & 34.6 & 38.9 & 28.7 \\
        Filled & 8.09 & 12.8 & 16.1 & 14.9 & 12.0 & 18.8 & --\\ \midrule \hline
        \multicolumn{8}{c}{\citet{ZHI201915} SBR} \\
        {} & Unaged & 1-day & 3-day & 5-day \\
        \midrule
        {} & 11.18 & 12.2 & 15.6 & 28.4  \\  \midrule \hline
        \multicolumn{8}{c}{\citet{Abdelaziz2021} SBR ($T=100^{\circ}C$)} \\
        Unaged & 7-day & 14-day & 21-day & 28-day  & 35-day & 45-day & 60-day\\
        \midrule
         11.2 & 16.5 & 22.2 & 32.0 & 29.1 & 22.4 & 27.1 & 20.6 \\ \midrule 
        \multicolumn{8}{c}{\citet{Abdelaziz2021} SBR ($T=90^{\circ}C$)} \\
        Unaged & 7-day & 14-day & 21-day & 28-day  & 35-day & 45-day & 60-day\\
        \midrule
         11.2 & 12.5 & 12.2 & 7.13 & 5.86 & 5.48 & 8.67 & 6.19 \\ \midrule
        \multicolumn{8}{c}{\citet{Abdelaziz2021} SBR ($T=70^{\circ}C$)} \\
        Unaged & 7-day & 14-day & 21-day & 28-day  & 35-day & 45-day & 60-day\\
        \midrule
         11.2 & 14.02 & 10.2 & 8.83 & 5.69 & 10.7 & 9.62 & 8.69 \\ \midrule
        \multicolumn{8}{c}{\citet{hamed1999tensile} NR} \\
        {} & Unaged & 4-day & 8-day & 10-day & 12-day \\
        \midrule
        Unfilled & 3.57 & 13.5 & 17.1 & 10.7 & 44.7 \\
        Filled & 9.82 & 7.46 & 7.97 & 12.2 & 13.9 \\  \bottomrule
    \end{tabular}
\label{Table: errors}
\end{table}

\section{Conclusions} \label{sec: Conclusions}
A physics-based and thermodynamically consistent constitutive equation for the thermo-chemical aging response of elastomers coupled with mechanical deformation has been proposed in this paper. The constitutive equation is based on modifying the Helmholtz free energy to include extra stored energy in the material upon aging. The effect of network crosslink reforming in modifying the shear modulus and chain segments is considered. The modification is based on chemical characterization tests, namely equilibrium swelling experiment, to measure the crosslink density evolution. The developed constitutive equation can predict the mechanical responses of thermo-chemically aged elastomers independent of mechanical tests on aged specimens within reasonable accuracy. The main contribution of this work is to connect the thermodynamic-based formulation and the form of stored energy directly to chemical characterization experiments. Unlike all previous studies, instead of conjecturing and assuming the form of energy storage, in this work, we develop a robust approach to obtain the stored energy as a function of crosslink formation. The proposed constitutive relationships thus provide a one-to-one mapping between chemically-based quantities (i.e., crosslink density) and physically-based macroscopic variables (i.e., shear modulus and chain segments). 

Comparison with experimental results showed that for the set of commercially available elastomers; more specifically filled and unfilled SBR and NR; the crosslink formation is the dominant network evolution. For the temperature range of up to 100$^{\circ}$C and up to 16 days, the additional stored energy based on the crosslink formation can accurately predict the thermo-chemo-mechanical responses of elastomers. By increasing the temperature above 100$^{\circ}$C, the rate of chain scission rapidly increases, and energy dissipation must be taken into account for accurate predictions. Moreover, the developed constitutive equation in this work can be easily incorporated into the more complicated diffusion-reaction-based constitutive equations to be fully coupled with the diffusion equations. The physically-based Helmholtz free energy proposed in this work can be used instead of phenomenologically assumed mechanical Helmholtz free energy in previous works, where the comprehensive reaction-driven Clausius-Duhem inequality were derived. This integration is the topic of future work by the authors.




\section*{Acknowledgement}
The authors gratefully acknowledge the support from the National Science Foundation under the award number CMMI-1914565.


\bibliography{ElastomerAging}



\end{document}